\documentclass[final,3p,times]{elsarticle}

\usepackage{graphicx}
\usepackage{amssymb}
\usepackage{amsmath}
\usepackage{comment}
\usepackage[colorlinks=true,linkcolor=NavyBlue,citecolor=NavyBlue,urlcolor=NavyBlue]{hyperref}

\biboptions{sort&compress} 

\makeatletter
\def\@author#1{\g@addto@macro\elsauthors{\normalsize%
    \def\baselinestretch{1}%
    \upshape\authorsep#1\unskip\textsuperscript{%
      \ifx\@fnmark\@empty\else\unskip\sep\@fnmark\let\sep=,\fi
      \ifx\@corref\@empty\else\unskip\sep\@corref\let\sep=,\fi
      }%
    \def\authorsep{\unskip,\space}%
    \global\let\@fnmark\@empty
    \global\let\@corref\@empty  
    \global\let\sep\@empty}%
    \@eadauthor={#1}
}
\makeatother

\makeatletter
\def\ps@pprintTitle{%
 \let\@oddhead\@empty 
 \let\@evenhead\@empty
 \def\@oddfoot{}%
 \let\@evenfoot\@oddfoot}
 \makeatother

\journal{arXiv}
\begin{document}
\begin{frontmatter}

\title{Secular bipolar growth rate of the real US GDP per capita: \\ 
implications for understanding past and future economic growth}

\author[add1,add2]{Sandro Claudio Lera}
\ead{slera@ethz.ch}

\author[add1,add3]{Didier Sornette}
\ead{dsornette@ethz.ch}

\address[add1]{ETH Zurich, Department of Management, Technology, and Economics, Scheuchzerstrasse 7, 8092 Zurich, Switzerland}
\address[add2]{ETH Zurich, Singapore-ETH Centre, 1 CREATE Way, \#06-01 CREATE Tower, 138602 Singapore}
\address[add3]{Swiss Finance Institute, c/o University of Geneva, Geneva, Switzerland}

\begin{abstract}
We present a quantitative characterisation of the fluctuations of
the annualized growth rate of the real US GDP per capita growth at many scales,
using a wavelet transform analysis of two data sets, quarterly data from 1947 to 2015
and annual data from 1800 to 2010. Our main finding is that 
the distribution of GDP growth rates can be well approximated by a bimodal function
associated to a series of switches between regimes of strong growth rate $\rho_\text{high}$
and regimes of low growth rate $\rho_\text{low}$. The succession
of such two regimes compounds to produce a
remarkably stable long term average real annualized growth rate of 1.6\% from 1800 to 2010 
and $\approx 2.0\%$ since 1950, which is the result of a subtle compensation 
between the high and low growth regimes that alternate continuously. Thus, the overall growth dynamics 
of the US economy is punctuated, with phases of strong growth that are intrinsically unsustainable, 
followed by corrections or consolidation until the next boom starts.  
We interpret these findings within the theory of ``social bubbles'' and
argue as a consequence that estimations of the cost of the 2008 crisis may be misleading.
We also interpret the absence of strong recovery since 2008 as a protracted
low growth regime  $\rho_\text{low}$ associated with the exceptional nature of the preceding 
large growth regime.
\end{abstract}

\begin{keyword}
Wavelet transform \sep Economic growth \sep GDP per capita \sep Business cycle
\PACS  89.65.Gh, 05.45.Tp
\end{keyword}
\end{frontmatter}

\section{Introduction}
\label{sec:Introduction}

The dynamics of the growth of GDP (gross domestic product), where GDP is defined as the market value of all final goods and services 
produced within a country in a given period of time \cite{Mankiw2011}, is arguably the most
scrutinised metric quantifying the overall economic development of an economy. A weak
annual growth rate of GDP, as has been characterising the US and Europe in the years following the
financial crisis of 2008, is interpreted as underperformance, which has called 
for unorthodox monetary policies \cite{Erber2012} to attempt to fix it. In contrast, a strong growth of GDP is usually 
lauded, because it usually reflects a rise of living standards and is generally
accompanied by decreasing unemployment. But what is meant by ``weak'' or ``strong'' growth?
Is there a ``natural'' growth rate? Does past growth rates of GDP imply future growth rates?
This last question is particularly relevant in the present context of small growth compared
with previous decades in developed countries and the argument by many that we may have
shifted to a ``new normal'' of slower intrinsic growth \cite{Dabla-Norris2015}.

A number of authors, e.g.  \cite{Holden2012,Fernald2014}, have noted that a plot of the logarithm of the US GDP 
as a function of (linear) time over the last one hundred years looks remarkably linear, as
shown by the continuous line and its dashed linear fitted line in figure \ref{fig:GDP_heatmap}:
the inflation adjusted GDP per capita exhibits a long term average growth of
1.9-2\% per year \cite{Fernald2014}. The occurrence of 
such a near trend-stationary long run growth covering a period with two world wars, 
the cold war and its associated proxy wars, the collapse of the Bretton Woods System in 1973, several large bubbles, crashes
and recessions and strong changes in interest rate policies, is truly remarkable.
It entices one to entertain the possibility of an equilibrium or natural growth rate, which 
then could be extrapolated in the future. Gordon \cite{Gordon2012} questions this extrapolation on the basis
of the drags that are bound to impact growth, including demography, education, inequality,
globalization, energy/environment, and the overhang of consumer and government debts.
Fernald and Jones \cite{Fernald2014} point out the large uncertainties associated
with new technologies, inequality, climate change and the increasing shift of the economy
towards health care. Holden \cite{Holden2012} observes that there are large 
medium frequency fluctuations around this linear trend of 1.9-2\% per year and presents
a model in which standard business cycle shocks \cite{Hansen1985,Merz1995,Bernanke1999}
lead to highly persistent movements around the long-term trend without significantly altering the trend itself,
due to a quasi-cancellation between the changes of new products and of new firms as a function of time.

Because of the wide spread effects that business 
cycles have both in society and in fiscal policy making \cite{Barseghyan2013}, 
researchers have been urged to develop a more solid understanding of this stylized fact. 
In recent years, the existence of out of equilibrium business cycles has been gaining 
more acceptance in economic theory. 
It is now understood that (out of equilibrium) business cycles, i.e. (excessive) periodic fluctuations in productivity, 
have significant effects on the cost of external finance \cite{Mclean2014}, on inflation \cite{Altig2011}, 
on employment and on many other macroeconomic key indicators \cite{Michele2001}. 
Business cycle-related market volatility has been shown to have predictive power 
on expected market returns \cite{Kim2013}, which, in turn, play a central role in the capital asset pricing model
at the heart of finance.  Noted professionals \cite{Dalio2015} view GDP growth as a long term trend, overlaid with both 
short and long term debt cycles, suggesting that fluctuations associated with business cycles could occur at all scales. 

Here, we present a quantitative characterisation of the fluctuations of US GDP growth at all scales,
using the natural tool for this, namely the wavelet transform. Adapting the analysing tool 
to the quantification of local growth rates, our main finding is that 
the distribution of GDP growths can be well approximated by a bimodal function.
One can thus represent the growth of the US economy per capita as an alternation 
between regimes of strong growth rate $\rho_\text{high}$, associated with booms (or bubbles), 
and regimes of low growth rate $\rho_\text{low}$ that include plateaus (and recessions). 
These two types of regimes alternate, thus quantifying the business cycle and giving an effective long-term 
growth rate $\rho_\text{lt}$ that is between $\rho_\text{low}$ and $\rho_\text{high}$. 
While the existence of fluctuations around the long-term growth trend has been noted
by many others as mentioned above, to the best of our knowledge, this is the first time that it is shown that these
fluctuations can be classified in two broad families, suggesting two 
well-defined economic regimes that completely exhaust the possible phases.

The existence of a well-characterised strong growth regime with average growth rate $\rho_\text{high}$
often leads to the misleading expectations that it is the normal that reflects a well-functioning economy, while
the other mode of low growth $\rho_\text{low}$ is considered
abnormal, often interpreted as due to a surprising shock, bringing considerable dismay and pain, 
and leading to policy interventions. Our finding of a robust bimodal distribution of GDP growth rates
over the whole history of the US suggests that this interpretation is incorrect.
Rather than accepting the existence of the long-term growth rate as given, and interpreting the
deviations from it as perturbations, the bimodal view of the GDP growth suggests a completely
different picture. In this representation, the long-term growth rate is the result of a subtle
compensation between the high and low growth regimes that alternate continuously.
The overall growth dynamics that emerges is that the US economy is growing in a punctuated way \cite{Louzoun2003},
following phases of strong growth that are intrinsically unsustainable, followed by corrections or consolidation
until the next boom starts. In other words, the approximately long-term growth rate reflects an economy that
oscillates between booms and consolidation regimes. Because of the remarkable
recurrence of the strong regime and in view of its short-term beneficial effects, 
economists and policy makers are tempted (and actually incentivised) to form their expectations
based on it, possibly catalysing or even creating it in a self-fulfilling prophecy fashion even when the real productivity
gains are no more present, as occurred in the three decades before the 2008 crisis \cite{Sornette2014a}.

We suggest that the transient strong growth regimes can be rationalised within the framework
of the ``social bubble hypothesis'' \cite{Sornette2008,Gisler2009,Gisler2010,Gisler2011}, in the sense that they result from 
collective enthusiasm that are similar to those developing during financial bubbles, which foster collective attitude 
towards more risk taking, The social bubble hypothesis claims that strong social interactions 
between enthusiastic supporters weave a network of reinforcing feedbacks that lead to widespread endorsement and extraordinary commitment by those involved, beyond what would be rationalised by a standard cost-benefit analysis.
For a time, the economy grows faster than its long-term trend, due to a number of factors
that reinforce each other, leading to a phase of creative innovation (e.g. the internet dotcom bubble)
or credit based expansion (e.g. the house boom and financialisation of the decade before 2008). These regimes
then unavoidably metamorphose into a ``hangover'', the recovery and strengthening episode
until the next upsurge.
 
Performing a careful analysis at multiple scales and over different window sizes up to the largest one going back to 1800,
we also find that the long-term growth rate of real GDP per capita has actually not been perfectly constant, being lower at about
1.6\% from 1800 till the end of WWII and growing to 1.9-2\% from 1950 until 2007 and then slowing down to approximately
1.1\% over the last 8 years. Informed by the above proposition that the high growth regime has no reason
to be the norm, the slower growth since 2008 suggests for a different interpretation. Having exhausted
the measures that (somewhat artificially \cite{Sornette2014a}) boosted economic growth in the previous three 
decades before the 2008 crisis 
and notwithstanding the introduction of exceptional measures, broadly referred to as ``quantitative easing'', 
the innovations and productivity gains seem unable to return to those during ``thirty glorious years'' of 1950-1980, preventing 
the recurrence of the strong boom regimes with $\rho_\text{high}$, but rather remain in a protracted
low growth regime  $\rho_\text{low}$.

In the next section, we present the wavelet transform that we use to examine the GDP growth rate fluctuations 
over different time scales. In section \ref{sec:wavelet_analysis_GDP}, we present our results concerning the
analysis of US GDP data and section \ref{sec:conclusions} concludes.

\section{The wavelet transform}
\label{sec:WT}

Originally developed in geophysics as a mathematical tool to analyze seismic signals \cite{Morlet1982,Goupillaud1984}, the wavelet
transform has proven useful for data analysis in a variety of fields such as image processing \cite{Antonini1992}, astrophysics \cite{Slezak1990},  
turbulence \cite{Argoul1989} and generally whenever complicated interactions between events occurring at different scales appear \cite{Meyer1992}.

A $\psi$-wavelet transform $W_\psi$ is simply a projection of a signal $X(\tau)$ onto $t$-translated
and $s$-dilated versions of $\psi$ \cite{Goupillaud1984,Grossmann1984,Yiou2000}:
\begin{equation}
	W_\psi[X](s,t) = \int \limits_{-\infty}^{\infty} d\tau~ \psi \left(\tau-t; s \right) ~X(\tau). 
	\label{eq:WT_definition}
\end{equation}
We call $s$ the scale and $t$ the time parameter. The analyzing function $\psi$, called the wavelet, has to be a localized function 
both in time and frequency domain. Depending on the application, the wavelets must be endowed with several additional properties, 
see \cite{Daubechies1990,Daubechies1992,Debnath2002} for mathematical details. For our purposes, it is important for the wavelet to be properly normalized. 
Assuming that $\psi(t;s)$ is approximately zero for values of $t$ outside the interval $[-s,s]$, the wavelet transform has then 
the following intuitive interpretation: $W_\psi[X](s,t)$ is the weighted average of $X$ over the interval $[t-s,t+s]$. The
wavelet transform can thus be seen as a `mathematical microscope' that resolves local structures of $X$ at `position' (time) $t$ and at a `magnification' (scale) $s$. 
Denoting by $\ast$ the convolution operator, expression \eqref{eq:WT_definition} can also be written compactly as $W_\psi[X](s,t) = [X(\tau) \ast \psi(\tau;s)](t)$,
or, for brevity, just $X \ast \psi$. 

Replacing $\psi$ in \eqref{eq:WT_definition} by its $n$-th derivative $\psi^{(n)}$ 
corresponds to a $\psi$-analysis of the $n$-th derivative
of the time series $X(t)$ (up to a normalization factor), as a simple integration by parts derivation shows. In this context, $\psi = \psi^{(0)}$ is
also called the mother wavelet. Since the overall statistical characterization of complex structures depends only weakly on the choice of the mother wavelet \cite{Arneodo1993}, we will show here only results for the Gaussian mother wavelet $\psi(t;s) = \exp(- t^2 / 2 s^2) / \sqrt{2 \pi} s$.  We have checked that other real-valued mother wavelets give similar results. 

In this article, we use the wavelet transform to quantify 
the pattern of local slopes (giving the local growth rates) of the analyzed time series (logarithm of the real US GDP per capita). 
This amounts to replacing $\psi$ in \eqref{eq:WT_definition} by 
the first derivative $\psi^{(1)}$ of the Gaussian mother wavelet, up to a normalization. The normalization 
is chosen such that the wavelet transform of the test signal $X(t) = p t$ with constant slope $p$ 
gives exactly its slope $p$ for all times $t$ and all scales $s$. This leads to the following expression for our analyzing mother wavelet used
in expression in \eqref{eq:WT_definition}:
\begin{equation}
	\psi^{(1)}(t;s) = \frac{t}{\sqrt{2 \pi} s^3} \exp \left( - \frac{1}{2} \left( \frac{t}{s} \right)^2 \right). 
	\label{eq:psi(1)}
\end{equation}
Note also that, by construction, the wavelet transform performed with $\psi^{(1)}(t;s)$ of a constant signal is zero.
This means that our implementation of the wavelet transform \eqref{eq:WT_definition} with \eqref{eq:psi(1)}
is insensitive to the absolute level and only quantifies precisely the local slope at a scale $s$.

In the remainder of this article, all figures are the result of the wavelet transform $X \ast \psi^{(1)}$ with $\psi^{(1)}$ given by \eqref{eq:psi(1)}.

\section{Wavelet analysis of the growth of real US GDP per capita}
\label{sec:wavelet_analysis_GDP}

\subsection{Analysed data: why real US GDP per capita?}

In this section, we analyze the real, i.e. adjusted for inflation, US GDP per capita (r-US-GDP-pc) as a measure for real 
innovation and productivity gains. In contrast, the total nominal US GDP contains two additional 
contributions to its growth: (i) population growth, including immigrants who are still an important component in the US;
(ii) inflation. In other words, the observed total US GDP can grow just from population change and/or the existence of inflation, 
even at constant or decreasing real US GDP per capita, often leading to a mistaken illusion of improved wealth \cite{Miao2013}.
Given that the flow of immigrants and population increase has undergone many complex phases
during the history of the US, it is important to disentangle this population component of GDP growth and 
work in units per capita. Similarly, inflation has been varying also significantly with bursts associated with wars,
the demise of the Bretton Woods System in 1973, the oil shocks and other financial crises.
The real US GDP per capita (r-US-GDP-pc) thus constitutes the standard gauge to evaluate 
the evolution over time of the real wealth per person, i.e. the one that counts, is felt in real terms
and is truly associated with progress.
We thus analyze the quarterly r-US-GDP-pc data between 1947 and 2015. 
In \ref{sec:annual_figures}, we present a similar analysis for the annual r-GDP-pc data 
over the much larger extended period between 1800 and 2010.

\subsection{Hierarchical structure of the GDP growth rates revealed by wavelet analysis}

\begin{figure}[!htb]
	\centering
	\includegraphics[width = \textwidth]{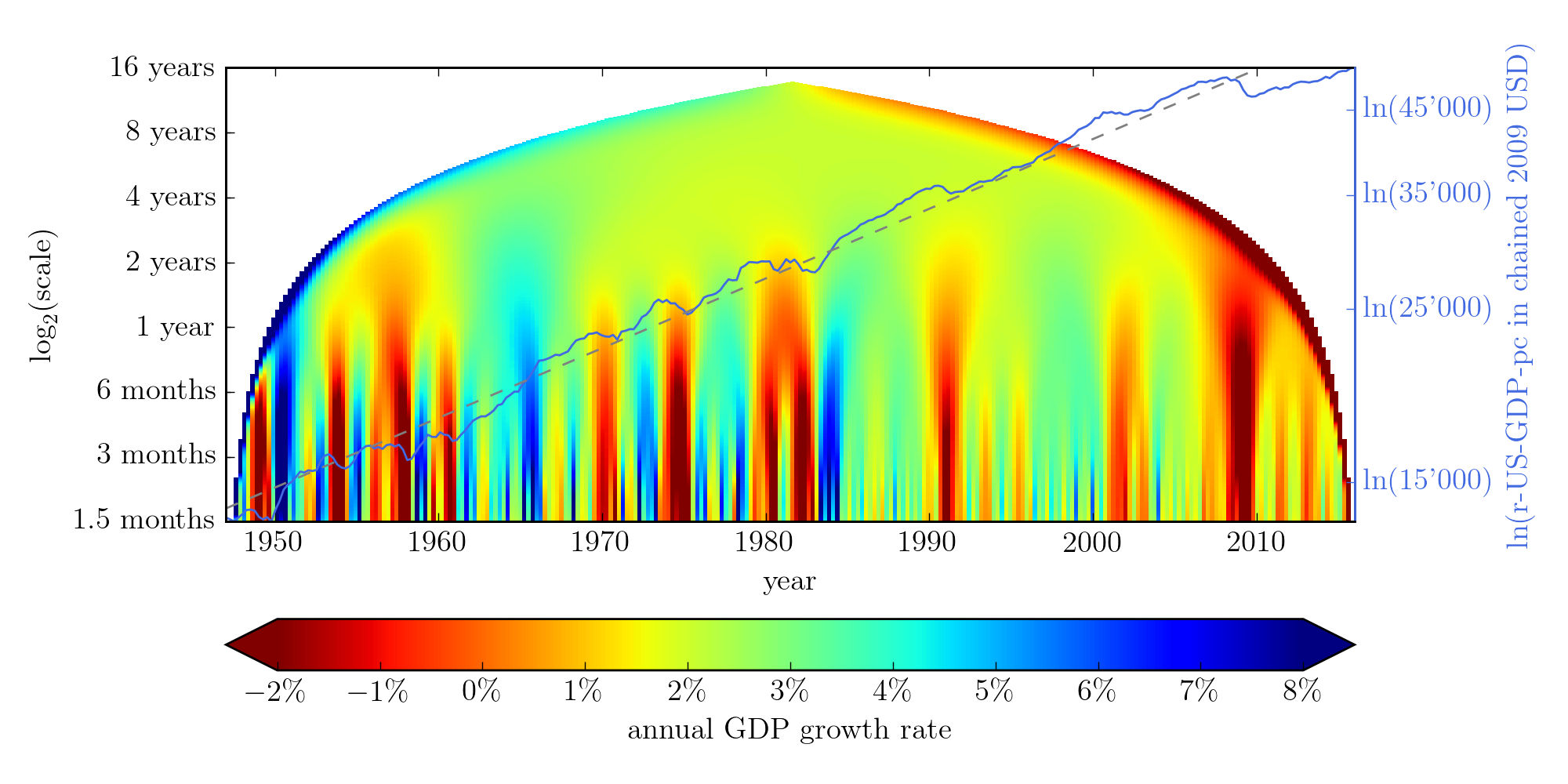}
	\caption{Wavelet transform $\ln(\text{r-US-GDP-pc}) \ast \psi^{(1)}$ of the logarithm of the quarterly real US GDP per capita data measured in chained 2009 US dollar over the period from 1947 to 2015 and represented by the continuous dark line (right vertical axis). An ordinary least squares fit determines a long-term annualized growth rate $\rho_\text{lt}$ of approximately $2\%$, shown as the dashed line. 
The left vertical axis plots the scale $s$ of the wavelet analysis, corresponding
to an interval of analysis approximately equal to $2s$. The color scale encodes the value of the annualized growth rates at different times and 
scales. The nonlinear conical shape of the envelop is due to edge-effects in the wavelet transform.}
	\label{fig:GDP_heatmap}
\end{figure}

As mentioned in the introduction, plotting the r-US-GDP-pc in a semi-logarithmic plot (figure \ref{fig:GDP_heatmap})
shows, to a first approximation, a remarkably straight line, suggesting that the r-US-GDP-pc 
grows exponentially as $\exp( \rho_\text{lt} t )$ with $t$ in units of years and 
a long-term annual growth rate $\rho_\text{lt} \approx 2\%$ determined by an ordinary least squares (OLS) fit.
This value is often reported in the literature as the average long-term historical growth of real GDP per capita (e.g. \cite{Fernald2014}). 

Beyond this long term average growth, one can see deviations that occur again and again.
Moreover, it is interesting to observe that the long-term growth rate $\rho_\text{lt}$ represented
by the slope of the straight dashed line seems to almost never describe the actual local growth rate of the r-US-GDP-pc.
In other words, the average growth rate does not seem to be a good description of the typical growth rates.
To quantify these qualitative observations, we perform a wavelet transform analysis of the logarithm of the the r-US-GDP-pc
at different times $t$ and different scales $s$ to obtain the local growth rate at time $t$, averaged over a time interval $[t-s,t+s]$,
defined by
\begin{equation}
\rho(s,t) = \ln(\text{r-US-GDP-pc}) \ast \psi^{(1)}~.
\end{equation}
The results are encoded with the color scale for the annualized growth rates in figure \ref{fig:GDP_heatmap} 
over the period from 1947 to 2015 shown on the horizontal axis. The left vertical axis plots the scale $s$ of the wavelet analysis, corresponding
to an interval of analysis approximately equal to $2s$. For scales at and lower than $s \approx 4$ years 
(i.e. averaged over approximately 8 years), one can first observe a hierarchy of branches 
associated with alternating warm (low or negative growth rates) and cold (positive and strong growth rates) colors.
As one goes to smaller and smaller time scales, more fine structures of alternating colors (growth rates) can be seen.
At the larger scales, $s \geqslant 4$ years, the color settles to the green value, recovering the known, and also directly determined by OLS, 
 long term growth $\rho_\text{lt} \approx 2\%$. 

Because the continuous wavelet transform \eqref{eq:WT_definition} with \eqref{eq:psi(1)} 
contains a lot of redundant information (a function $X(t)$ of one variable $t$
is transformed into a function $W_\psi[X](s,t)$ of two variables $t$ and $s$), it is standard to compress
the wavelet map shown in figure \ref{fig:GDP_heatmap} 
into a so-called 
``skeleton'' \cite{Arneodo2002,Mallat1992a,Mallat1992b}. The skeleton of $W_\psi[X](s,t)$
is the set of all local maxima of $\left| W_\psi[X](s,t) \right|$ considered as a function of $t$, for fixed scale $s$. 
It is thus the set of all local maxima and minima of $W_\psi[X](s,t)$.
The skeleton forms a set of connected curves in the time-scale space, called the extrema lines. 
Geometrically, each such skeleton line corresponds to either a crest or valley bottom of the three-dimensional 
representation of the wavelet function $W_\psi[X](s,t)$. A crest can be viewed as the typical value
of the growth rate of a locally surging r-US-GDP-pc. The bottom of a valley is similarly the typical value
of the growth rate of a locally slowing down or contracting r-US-GDP-pc.

\begin{figure}[!htb]
	\centering
	\includegraphics[width = \textwidth]{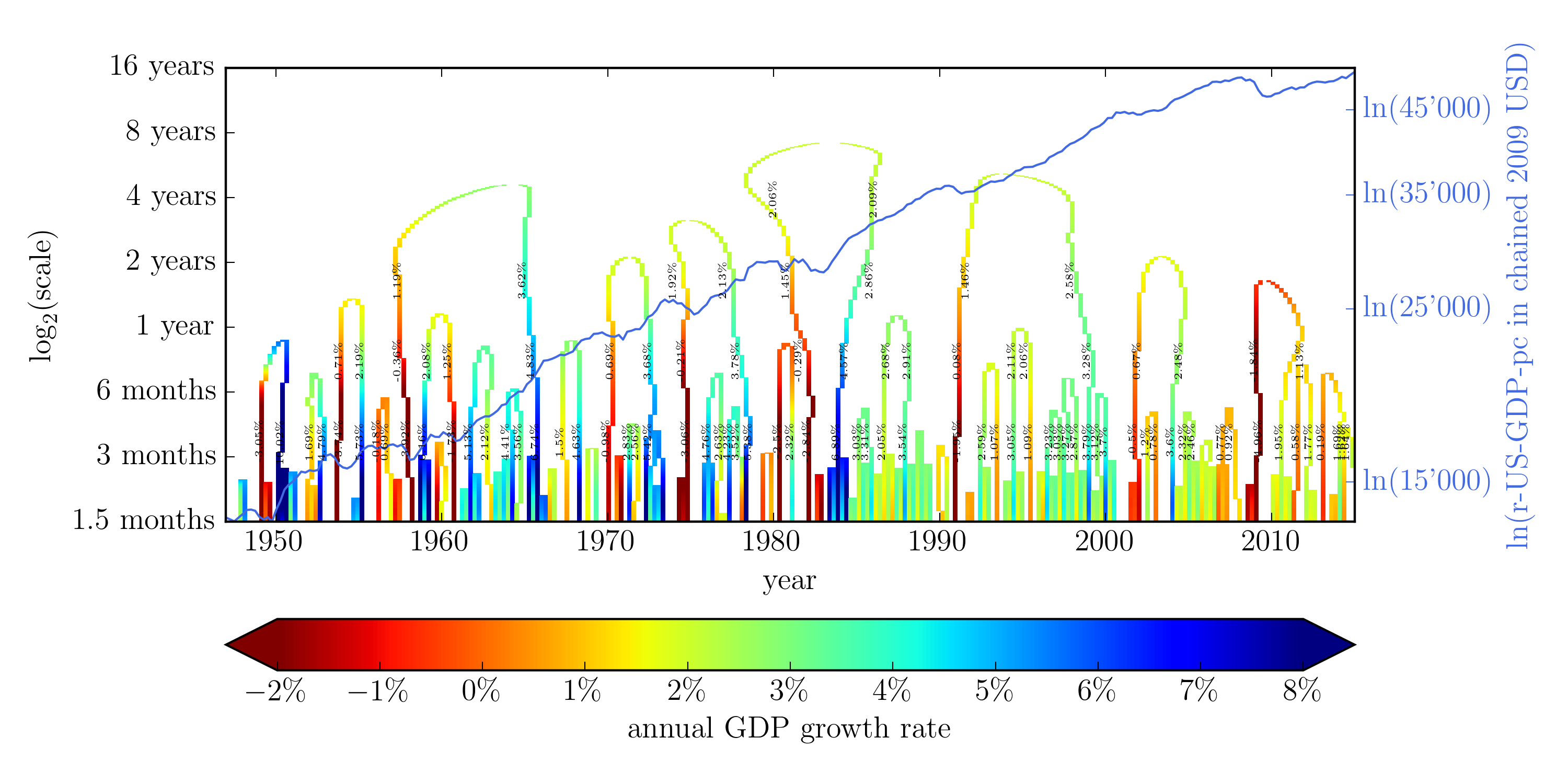}
	\caption{Skeleton structure of the wavelet transform $\ln(\text{r-GDP-pc}) \ast \psi^{(1)}$ for quarterly real US GDP per capita data measured in chained 2009 US dollar corresponding to figure \ref{fig:GDP_heatmap}.}
	\label{fig:GDP_skeleton}
\end{figure}

The skeleton of figure \ref{fig:GDP_heatmap}
is shown in figure \ref{fig:GDP_skeleton}. 
One can observe
much more clearly the hierarchy of alternating growth regimes, which combine into an overall growth of $\approx 2\%$ at large scales. 
Written along each skeleton line in the figure, we give the values of the local annualized growth rates 
at four scale levels, 3 months, 6 months, 18 months
and 3 years. The structure of the skeleton lines, their colors and the values of the local annualized growth rates
confirm the existence of ubiquitous shifting regimes of slow and strong growths. 

To validate the intuition that a crest (resp. valley bottom) of the skeleton can be viewed as the typical value
of the growth rate of a locally surging (resp. slowing down) r-US-GDP-pc, we check that 
only knowing the growth rates on the skeleton structure retains the structural information of the full wavelet transform.
For this, we have constructed synthetic versions of the real GDP per capita 
from the skeleton at different scales as explained in \ref{sec:pseudo_GDP}, and we have compared them
to the true r-US-GDP-pc.

\subsection{Evidence for a robust bimodal structure of distributions of US GDP growth rates}

\begin{figure}[!htb]
	\centering
	\includegraphics[width = \textwidth]{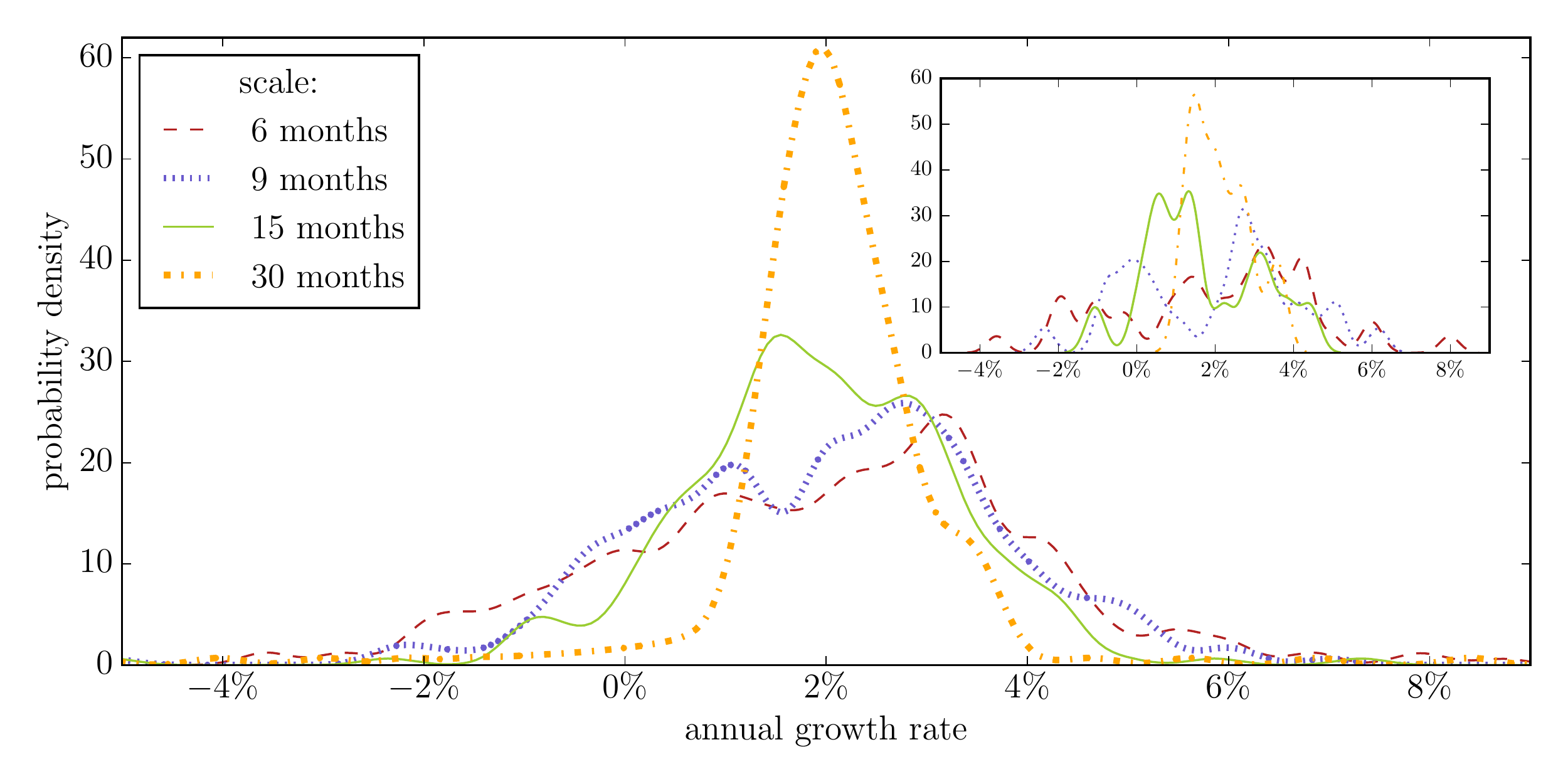}
	\caption{Gaussian kernel estimations (with width equal to $0.002$) of the probability density distributions (pdf) of the
	local annualized growth rates of r-US-GDP-pc at four different scales indicated in the inset in the top-left. 
	 The main panel represents the distributions extracted from the wavelet transform 
	shown in figure \ref{fig:GDP_heatmap}, while the top-right inset shows the pdf's obtained from the skeleton values
	shown in figure \ref{fig:GDP_skeleton}.}
	\label{fig:GDP_distributions}
\end{figure}

The nature of the shifting regimes of slow and strong growths
can be quantified further by constructing the probability density distributions (pdf) of annualized 
GDP growth rates at different fixed scales, 
both from the entire wavelet transform (figure \ref{fig:GDP_heatmap})
and from the skeleton structure (figure \ref{fig:GDP_skeleton}). 
The obtained pdf's for four different scales (6, 9, 15 and 30 months) are depicted in figure \ref{fig:GDP_distributions}. 
They have been obtained using a Gaussian kernel estimations with width equal to $0.002$.
We have checked the robustness of these pdf's by changing the width of the kernels within a factor of two.

The pdf's extracted from the wavelet transform shown in figure \ref{fig:GDP_heatmap}
and from the skeleton values shown in figure \ref{fig:GDP_skeleton} exhibit the same structures.
First, the pdf's at the largest scale of 30 months peak at the annualized growth rate of $\approx 2\%$,
recovering the OLS value reported above (shown as the dashed line in figure \ref{fig:GDP_heatmap}).
Second, as we go down to smaller scales, already at the scale of 15 months, and more pronounced
at the scale of 9 and 6 months, a clear bimodal structure emerges (decorated by higher frequency structures,
associated with the width of the estimating kernel). 
Denoting the two main peaks of the bimodal density extracted from the full wavelet transform (the skeleton gives similar results) 
at scale $s$ by $\rho_\text{low}(s)$ and $\rho_\text{high}(s)$ respectively,  we obtain
\begin{equation}
\rho_\text{low}(6 \text{ months}) \approx 1\% \lesssim
\rho_\text{low}(9 \text{ months}) \approx 1.1\% \lesssim 
\rho_\text{low}(15 \text{ months}) \approx 1.5\% \lesssim
\rho_\text{lt} \approx 2\%
\end{equation}
and 
\begin{equation}
\rho_\text{high}(6 \text{ months}) \approx 3.1\% \gtrsim
\rho_\text{high}(9 \text{ months}) \approx 2.8\% \approx
\rho_\text{high}(15 \text{ months}) \approx 2.8\% \gtrsim
\rho_\text{lt} \approx 2\%. 
\end{equation}
The pleasant stability for the estimates $\rho_\text{low}(6 \text{ months}) \approx \rho_\text{low}(9 \text{ months})$
and $\rho_\text{high}(6 \text{ months}) \approx \rho_\text{high}(9 \text{ months}) \approx \rho_\text{high}(15 \text{ months})$
suggests that real US GDP per capita can be modelled as an alternation of slow growth around a typical value 
of $1\%$ and strong growth around a typical value of $3\%$, which bracket the long-term average growth rate 
$\rho_\text{lt} \approx 2\%$. This constitutes the main result of our article.

\ref{sec:annual_figures} presents the wavelet transform, skeleton structure and growth rate
distributions for annual r-US-GDP-pc data starting in 1800 till 2010.  The important conclusion is that 
the previous observations presented above for quarterly data from 1950 to 2015 are broadly confirmed
when using annual data over this much longer period.

\section{Concluding remarks}
\label{sec:conclusions}

We have presented a quantitative characterisation of the fluctuations of
the annualized growth rate of the real US GDP per capita growth at many scales,
using a wavelet transform analysis of two data sets, quarterly data from 1947 to 2015
and annual data from 1800 to 2010. Our main finding is that 
the distribution of GDP growth rates can be well approximated by a bimodal function
associated to a series of switches between regimes of strong growth rate $\rho_\text{high}$
and regimes of low growth rate $\rho_\text{low}$. The succession
of alternations of these two regimes compounds to produce a
remarkably stable long term average real annualized growth rate of 1.6\% from 1800 to 2010 and $\approx 2.0\%$ since 1950.

We thus infer that the robust constant growth rate since 1950 cannot be taken as
evidence for a purely exogenous ``natural'' rate of innovations and productivity growth.
It is rather a remarkable output of the succession of booms and corrections that punctuate
the economic history of the US since more than 200 years. 
Our results suggest that alternating growth regimes
are intrinsic to the dynamics of the US economy and appear at all scales. These alternating
regimes can be identified as generalized business cycles, occurring at the
scale of the whole economy. 

Such business cycles may be briefly rationalised as follows.
During the high growth regime, a number of positive feedback loops are in operation, such 
as deregulation, enhanced credit creation, the belief in a ``new economy" and so on. This creates a transient boom, 
perhaps accelerating itself and leading to financial and social bubbles 
\cite{Geraskin2013,Sornette2014,Yukalov2015,Sornette2008,Gisler2009,Gisler2011}.
This over heating of the economy then turns out not to be sustainable and leads to a correction
and consolidation phase, the low growth regime. Then, the next strong growth regime starts
and so on. 

Our findings suggest that strong growth cannot be dissociated from periods of 
recessions, corrections or plateaus, which serve as a consolidation phase before
the next boom starts. However, because of the remarkable
recurrence of the strong regime and in view of its short-term beneficial effects, 
economists and policy makers tend to form expectations of strong continuous growth.
Such way of thinking may lead to conclusions that, we argue, may have little merit. Consider the estimation 
of the US Federal Reserve Bank of Dallas \cite{Atkinson2013} that 
the cost of the 2008 crisis, assuming output eventually returning to its pre-crisis trend path,
is an output loss of \$6 trillion to \$14 trillion US dollars. These enormous numbers are
based on the integration of the difference between the extrapolation of hypothetical 
GDP trajectories expected from a typical return to pre-crisis growth compared with the
realised GDP per capita. In the light of our findings, we argue that it is incorrect to 
extrapolate to the pre-crisis growth rate, which is by construction abnormally high, and
much higher than the long term growth rate. In addition, one should take into account the 
fact that the base rate after a crisis should be low or even negative, for the consolidation to work. 
Moreover, the duration of the boom years may have direct impact on that of the recovery period.
In this vein, Sornette and Cauwels \cite{Sornette2014a}
have argued that this 2008 crisis is special, as it is the culmination of a 30 year trend
of accelerating financialization, deregulation and debt growth. Our present results impel the 
careful thinker to ponder what is the ``natural'' growth rate and avoid naive extrapolations.

Using a simple generic agent-based model
of growth, Louzoun et al. \cite{Louzoun2003} have identified 
the existence of a trade off between either low and steady growth 
or large growth associated with high volatility. Translating this insight to the US economy
and combining with the reported empirical evidence, the observed growth 
features shown in the present paper seem to reveal a remarkable stable
relationship between growth and its fluctuations over many decades, if not centuries.
Perhaps, this is anchored in the political institutions as well as in the psychology
of policy makers and business leaders over the long term that transcend
the short-term vagaries of political power sharing and geopolitics. It is however
important to include in these considerations the fact that the US is unique compared with other developed countries, having 
benefitted enormously from the two world wars in particular (compared with the 
destruction of the UK and French empires and the demise of the economic dominance of european powers).

\bibliographystyle{unsrt}
\bibliography{bipolar_growth_bibliography}

\appendix
\section{Analysis of annual US GDP data}
\label{sec:annual_figures}

In this appendix, we present wavelet transform, skeleton structure and growth rate
distributions for annual r-US-GDP-pc data starting in 1800 till 2010.  The important conclusion is that 
the previous observations presented in the main text for quarterly data from 1950 to 2015 are broadly confirmed
when using annual data over this much longer period.

\begin{figure}[!htb]
	\centering
	\includegraphics[width = \textwidth]{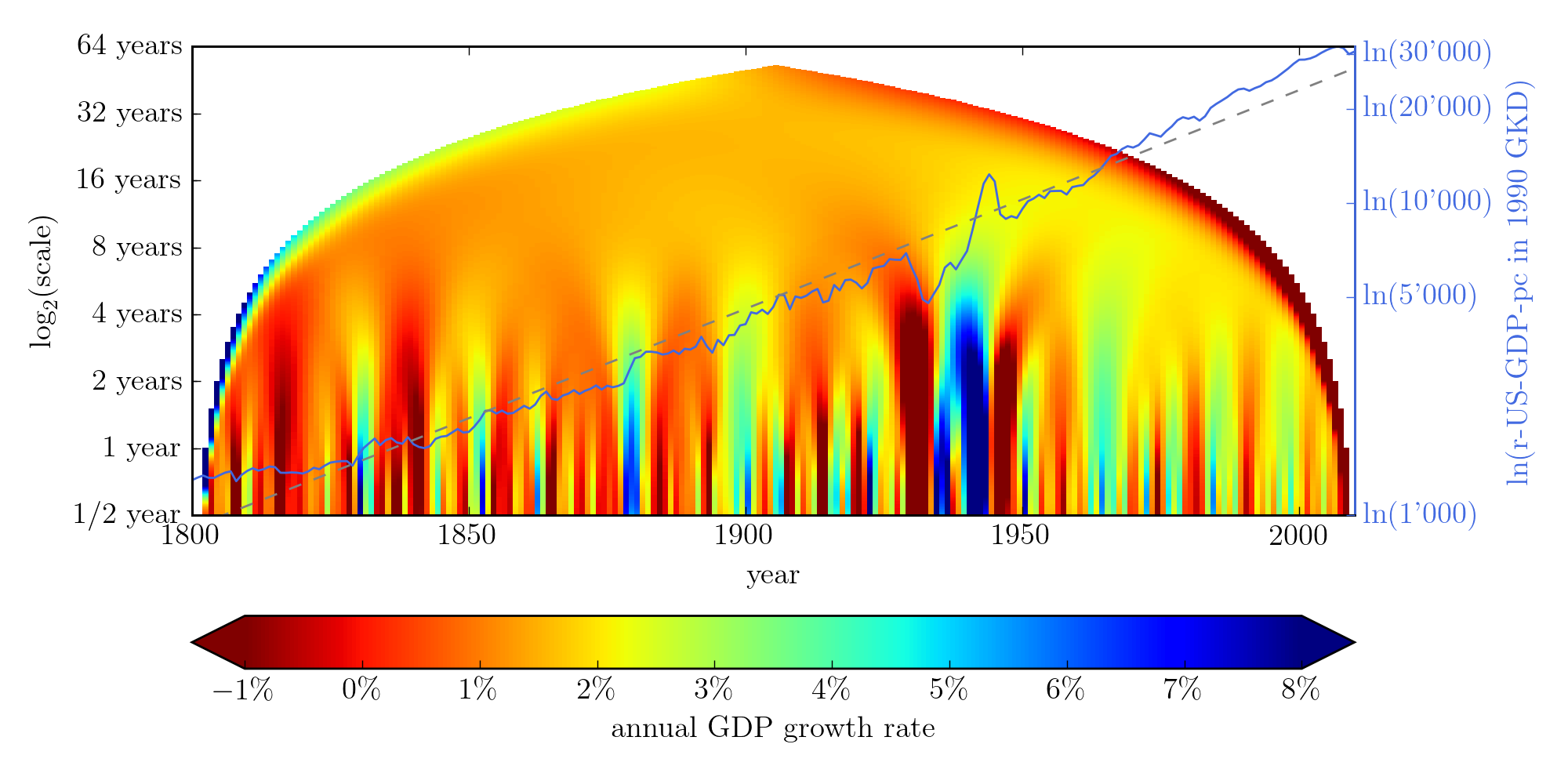}
	\caption{Wavelet transform $\ln(\text{r-US-GDP-pc}) \ast \psi^{(1)}$ of the logarithm of the annual real US GDP per capita data 
	measured in 1990 Geary-Khamis dollar and represented by the continuous dark line (right vertical axis). 
	An ordinary least squares fit determines a long-term annualized growth rate $\rho_\text{lt}$ of approximately $1.6\%$.}
	\label{fig:GDP_heatmap_annual}
\end{figure}

\begin{figure}[!htb]
	\centering
	\includegraphics[width = \textwidth]{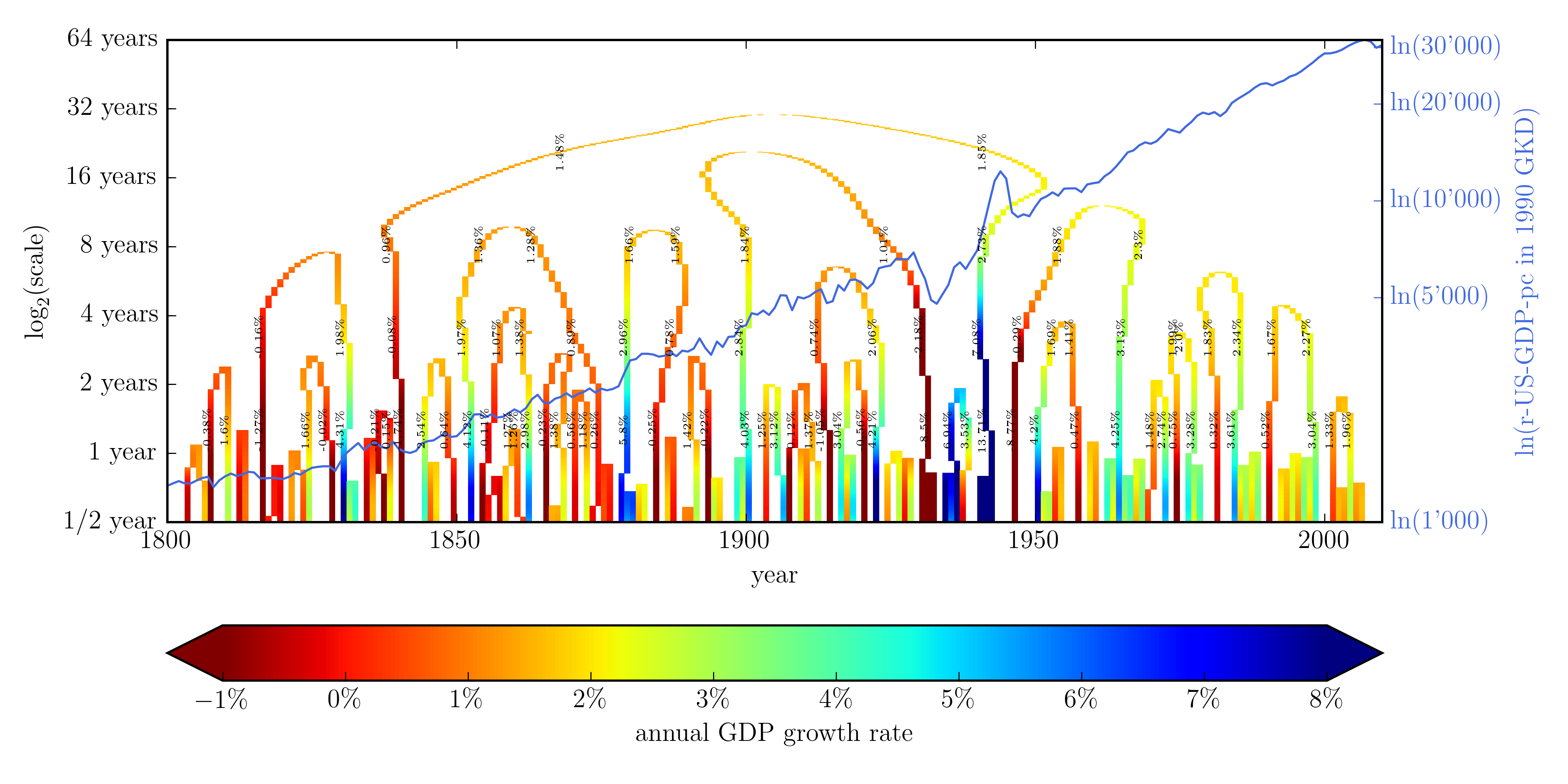}
	\caption{Skeleton structure of $\ln(\text{r-GDP-pc}) \ast \psi^{(1)}$  for annual real US GDP per capita data measured in 1990 Geary-Khamis dollar corresponding to figure \ref{fig:GDP_heatmap_annual}.}
	\label{fig:GDP_skeleton_annual}
\end{figure}

\begin{figure}[!htb]
	\centering
	\includegraphics[width = \textwidth]{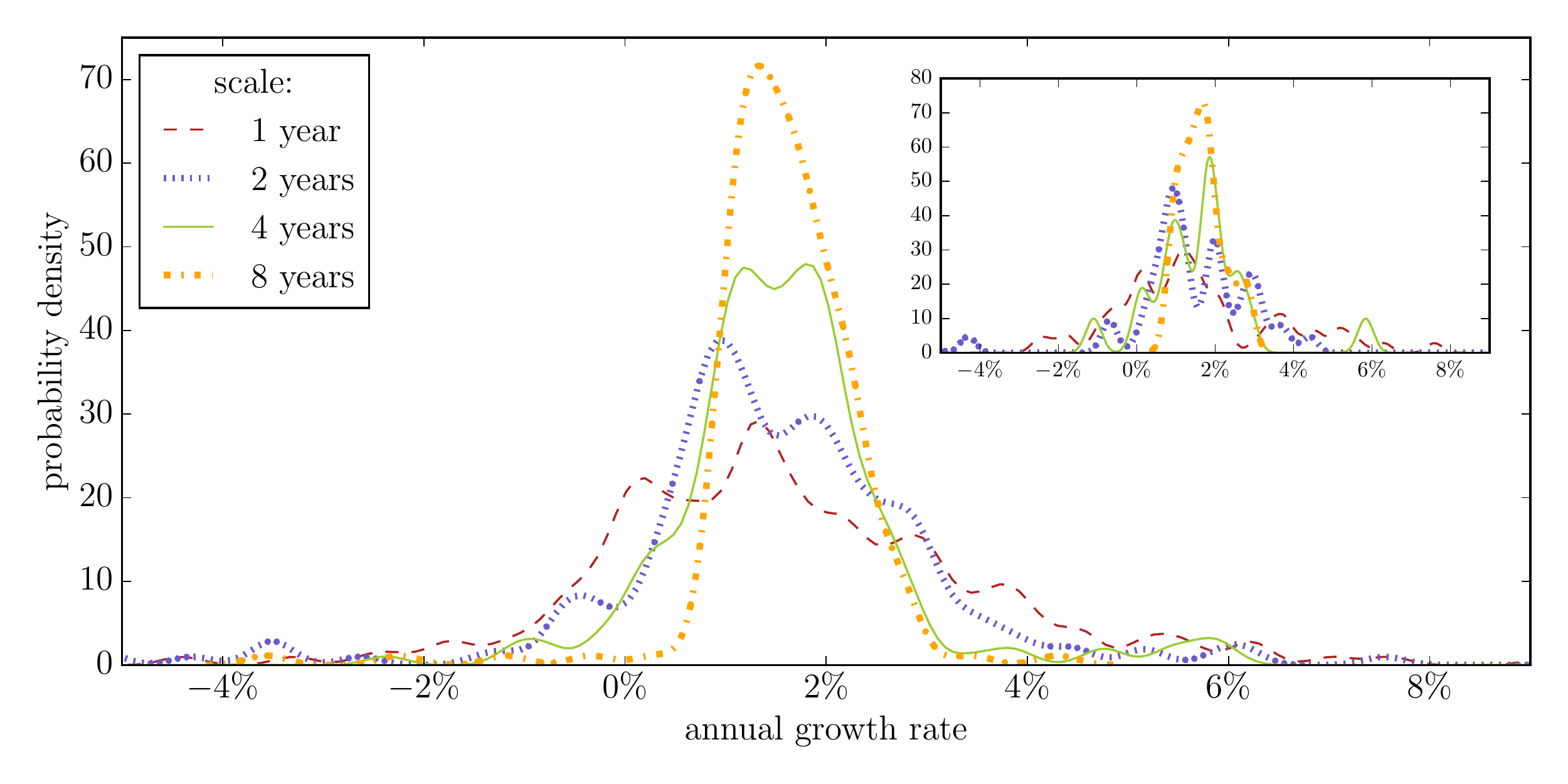}
	\caption{Gaussian kernel estimations (with width equal to $0.002$) of the probability density distributions (pdf) of the
	local annualized growth rates of r-US-GDP-pc sampled annually from 1800 to 2010 at four different scales indicated in the inset in the top-left. The main panel represents the distributions extracted from the wavelet transform 
	shown in figure \ref{fig:GDP_heatmap_annual}, while the top-right inset shows the pdf's obtained from the skeleton values
	shown in figure \ref{fig:GDP_skeleton_annual}.}
	\label{fig:GDP_distributions_annual}
\end{figure}

Figure \ref{fig:GDP_heatmap_annual} 
shows
the wavelet transform $\ln(\text{r-US-GDP-pc}) \ast \psi^{(1)}$ of the logarithm of the annual real US GDP per capita data 
measured in 1990 Geary-Khamis dollar from 1800 to 2010 and represented by the continuous dark line (right vertical axis). 
An ordinary least squares fit determines a long-term annualized growth rate $\rho_\text{lt}$ of approximately $1.6\%$.
This value is smaller than the average growth rate of $2\%$ determined for the period from 1950 to 2015.
This smaller value is a compromise, given the rather clear long term upward curvature presented by the continuous curve shown in 
figure \ref{fig:GDP_heatmap_annual}, expressing a tendency for the growth rate to grow itself \cite{Johanssen2001}.
Indeed, one can see that $ln(\text{r-US-GDP-pc})$ departs more and more from the dashed straight line
with a larger slope after 1950, in line with the observations shown in figure \ref{fig:GDP_heatmap}.
Figure \ref{fig:GDP_skeleton_annual} depicts the skeleton structure extracted from figure \ref{fig:GDP_heatmap_annual}
and figure \ref{fig:GDP_distributions_annual}
plots the distribution of growth rates
of the r-US-GDP-pc extracted from annual real US GDP per capita data since 1800. 
As for the quarterly GDP data, the long term growth rate of $\approx 1.5\%$ is recovered 
as the peak of the distribution at the $8$ year scale.
The bimodal structure is less clean, due to the fact that annual sampling of GDP growth rates
is bound to average over the time scales during which the transitions between 
the different regimes occur. Nevertheless, one can observe two main peaks,
except for the largest time scale of $8$ years. Moreover, both at the $1$ and $2$ year scales, 
the estimated probability density functions (pdf) exhibit a positive skewness, with an asymmetric
tail to the right side of large positive growth rates. This means that a considerable part 
of the probability mass is concentrated at high growth regimes above $\rho_\text{lt}$. Quantitatively, 
the values of the growth rates corresponding to the two main peaks of the pdf's are:
 $\rho_\text{low}(1,2,4 \text{ years}) = 0.2\%,1\%,1.2\%$ and 
$\rho_\text{high}(1,2,4 \text{years}) = 1.4\%, 1.9\%, 1.8\%$. The rather low value
$\rho_\text{high}(1 \text{year}) = 1.4\%$ should be considered together with the evidence
of the strong positive skewness noted above: at the annual sampling rate, the granularity of the 
data is too coarse to recover the clean picture of the quarterly data due to overlapping intervals.
However, we find that the $50\%$-quantile growth rate is $1.3\%$, while the expectation value  of the growth rates
conditional on growth rates above $1.3\%$ is equal to $3.3\%$, much larger than $\rho_\text{lt}$.
The results are thus in broad agreement with those presented  in figure \ref{fig:GDP_distributions}
for quarterly data.

\section{Synthetic GDP from wavelet skeletons}
\label{sec:pseudo_GDP}

\begin{figure}[!htb]
	\centering
	\includegraphics[width = \textwidth]{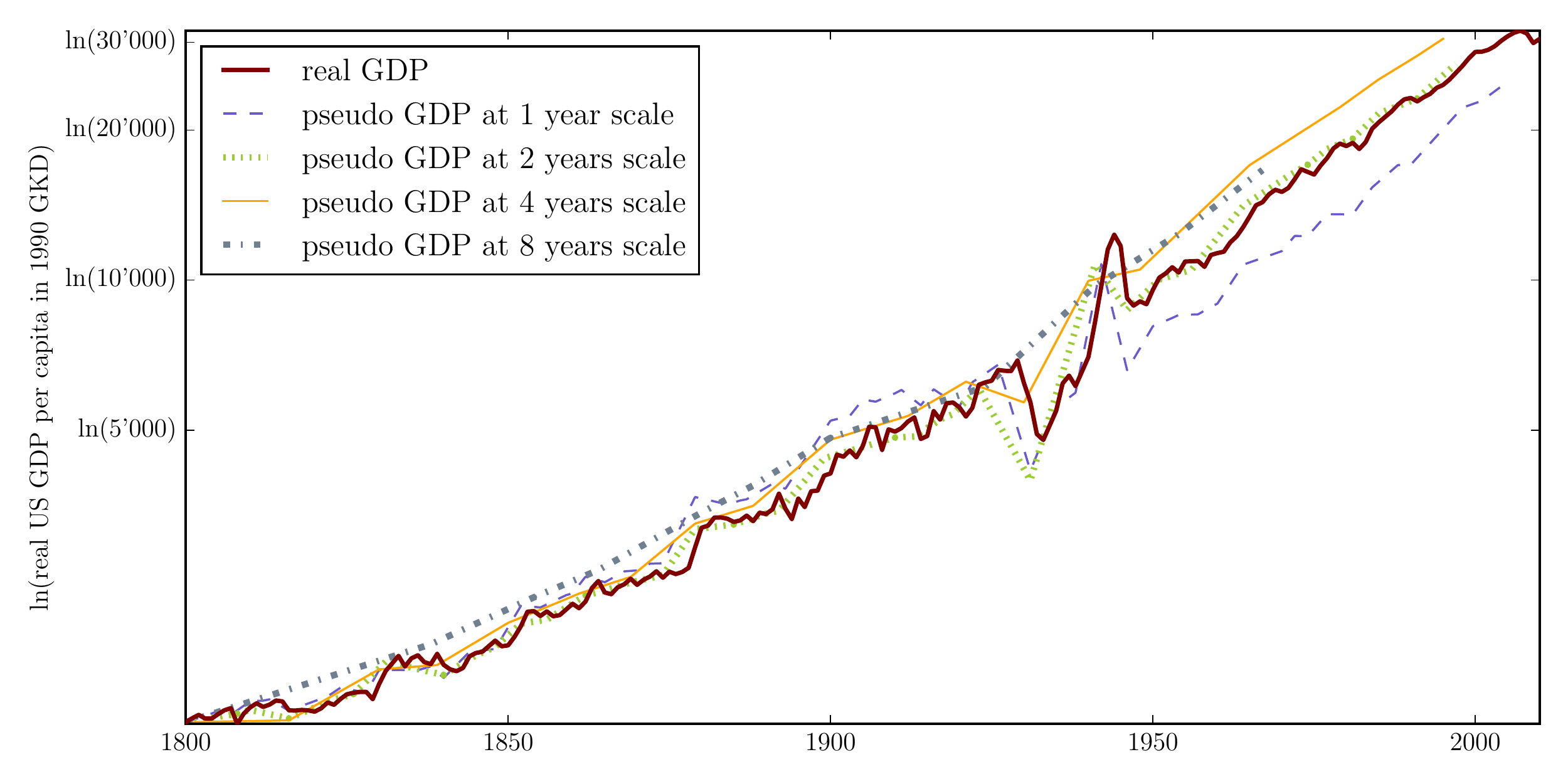}
	\caption{Synthetic real GDP per capita constructed using expression \eqref{eq:defi_pseudo_GDP} at four different scales and
	comparison with the realized real US GDP per capita.} 
	\label{fig:pseudo_annual}
\end{figure}

In figure \ref{fig:GDP_skeleton} and figure \ref{fig:GDP_skeleton_annual}, we have extracted the skeleton structure by keeping only 
local extrema as a function of time and discarding all additional information from the wavelet transform. 
In order to verify that the skeleton structure retains the most crucial information of the full wavelet transform, 
we construct here what we call the synthetic GDP, as follows: 
\begin{enumerate}  \itemsep0em
	\item Think of the skeleton structure of the real US GDP per capita (r-US-GDP-pc) growth rates shown in figure \ref{fig:GDP_skeleton_annual}. 

	\item Fix a scale $s^*$. 
	
	\item Imagine a horizontal line at that fixed scale $s^*$. This line intercepts some $n = n(s^*)$ skeleton arms of the wavelet
		transform. Denote the wavelet coefficient at the intercept of the horizontal line with 
		the $i$-th arm by $g_i$. This results in a set of $n$ growth rates $g_1, \ldots, g_n$. 
		
	\item Denote the time coordinate of the $g_i$ intercept by $t_i$. This way, we can
		uniquely label each $g_i$ as $g_i = g_i(t_i;s^*)$.
				
	\item Denote the temporal separation between $g_{i-1}$ and $g_i$ by $\Delta t_i \equiv t_i-t_{i-1}$ (in units of years).  
		
	\item Denote the origin of time by $t_0$. In our annual r-US-GDP-pc data set, $t_0 = 1800$.  
	
	\item Define $\ln(\text{GDP}_0) \equiv \ln \left( \text{r-US-GDP-pc in 1800} \right)$. 
		
	\item Then, the synthetic GDP at scale $s^*$ at time $t_k$ is calculated as 
		\begin{equation}
			 \text{GDP}_\text{pseudo}(t_k;s^*) \equiv \text{GDP}_0  \cdot \prod_{i=1}^k (1+g_i)^{\Delta t_i}. 
			\label{eq:defi_pseudo_GDP}
		\end{equation}
		
	\item We expect $\text{GDP}_\text{pseudo}(t_k)~ \approx ~$r-US-GDP-pc at time $t_k$ and to be 
	approximately independent of the scale $s^*$. 
\end{enumerate}
We have calculated the synthetic GDP given by formula \eqref{eq:defi_pseudo_GDP} at 
four different scales $s^* = 1$ year, $2$ years, $4$ years and $8$ years. 
The result is depicted in figure \ref{fig:pseudo_annual}. 
We see that the skeleton structure reproduces the real GDP quite accurately, thus confirming that the skeleton structure retains the most crucial features of the full wavelet structure. Notable deviations are observable around the sequence of peaks around the 1950s, where the discreteness of the skeleton structure leads to an overshooting or undershooting in the very unstable time associated with 
WWII and its aftermath, when the US GDP shot up to extraordinary level corresponding to the an outstanding war effort,
following by a crash back to the long term trend. The distorsion to the long-term synthetic GDP brought by this period
does not impact our study of the local growth rates and their bimodal distribution, being local in nature.

\end{document}